\documentclass[aps,prd,showpacs,twocolumn,groupedaddress]{revtex4}
\usepackage{graphicx}
\usepackage{amsfonts}
\usepackage{amssymb}
\usepackage{amsmath}
\def\lsim {~^{<~}_{\sim~}}
\def\gsim {~^{>~}_{\sim~}}

\usepackage{ulem}
\usepackage{color}

\begin{document}
\title{Off-diagonal Gluon Mass Generation and Infrared Abelian Dominance\\
in Maximally Abelian Gauge in SU(3) Lattice QCD}

\author{Shinya~Gongyo}
  \email{gongyo@ruby.scphys.kyoto-u.ac.jp}
  \affiliation{Department of Physics, Graduate School of Science,
  Kyoto University, \\
  Kitashirakawa-oiwake, Sakyo, Kyoto 606-8502, Japan}
\author{Takumi~Iritani}
  \email{iritani@ruby.scphys.kyoto-u.ac.jp}
  \affiliation{Department of Physics, Graduate School of Science,
  Kyoto University, \\
  Kitashirakawa-oiwake, Sakyo, Kyoto 606-8502, Japan}
\author{Hideo~Suganuma}
  \email{suganuma@ruby.scphys.kyoto-u.ac.jp}
  \affiliation{Department of Physics, Graduate School of Science,
  Kyoto University, \\
  Kitashirakawa-oiwake, Sakyo, Kyoto 606-8502, Japan}

\date{\today}
\begin{abstract}
In SU(3) lattice QCD formalism, we propose a method to extract gauge fields from link-variables analytically. With this method, we perform the first study on effective mass generation of off-diagonal gluons and infrared Abelian dominance in the maximally Abelian (MA) gauge in the SU(3) case. Using SU(3) lattice QCD, we investigate the propagator and the effective mass of the gluon fields in the MA gauge with U(1)$_3 \times$U(1)$_8$ Landau gauge fixing. The Monte Carlo simulation is performed on $16^4$ at $\beta$=5.7, 5.8 and 6.0 at the quenched level. The off-diagonal gluons behave as  massive vector bosons 
with the approximate effective mass $M_{\mathrm{off}} \simeq 1.1-1.2\mathrm{GeV}$ in the region of 
$r =0.3-0.8$fm, and the propagation is limited within a short range, while the propagation of diagonal gluons remains even in a large range.   
In this way, infrared Abelian dominance is shown in terms of 
short-range propagation of off-diagonal gluons.
Furthermore, we investigate the functional form of the off-diagonal gluon propagator. The functional form is well described by the four-dimensional Euclidean Yukawa-type function $e^{-m_{\rm off}r}/r$ with $m_{\rm off} \simeq 1.3-1.4\mathrm{GeV}$ for $r = 0.1- 0.8$ fm. This also indicates that the spectral function of off-diagonal gluons has the negative-value region.  
\end{abstract}
\pacs{12.38.Aw, 12.38.Gc, 14.70.Dj}
\maketitle

\section{Introduction}

Quantum chromodynamics (QCD) is the fundamental gauge theory of the strong interaction based on quarks and gluons. There are a variety of nonperturbative phenomena in low energy QCD such as color confinement and chiral symmetry breaking. These nonperturbative phenomena have been studied both in analytical frameworks and in lattice QCD \cite{C7980,R05,Cr11}.

On the quark-confinement mechanism, Nambu, 't Hooft and Mandelstam suggested the dual-superconductor picture \cite{N74}. This picture is based on the electromagnetic duality and the analogy with the one-dimensional squeezing of the magnetic flux in the type-II superconductor. In this picture, there occurs color magnetic monopole condensation, and then the color-electric flux between the quark and the antiquark is squeezed as a one-dimensional tube due to the dual Higgs mechanism.  From the viewpoint of the dual-superconductor picture in QCD, however, there are two assumptions of Abelian dominance \cite{tH81,EI82} and monopole condensation.
Here, Abelian dominance means that only the diagonal gluon component plays the dominant role for the nonperturbative QCD phenomena like confinement.

The maximally abelian (MA) gauge has mainly been investigated from the viewpoint of the dual-superconductor picture \cite{AS98,BC03,SST95,Ko11,SY90,IH99,Mi95,Wo95,KSW87,BWS91,SNW94,SAI02,HSA10} and the various lattice QCD Monte Carlo simulations show that the MA gauge fixing seems to support these assumptions \cite{AS98,BC03, SY90, IH99,Mi95, Wo95, KSW87, BWS91, SNW94,SAI02}.

According to these studies, the diagonal gluons seem to be significant to the infrared QCD physics, which is called ``infrared Abelian dominance". Infrared Abelian dominance means that off-diagonal gluons do not contribute to infrared QCD. Therefore, the essence of infrared Abelian dominance is the behavior of  the off-diagonal gluon propagator.

The gluon propagators in the MA gauge has been investigated in SU(2) lattice Monte Carlo simulations \cite{AS98, BC03,Cu01}.
 To investigate the gluon propagators in the MA gauge, 
it is desired to extract the gluons exactly from the link-variables, because the link-variable cannot be expanded even for a small lattice spacing due to the large fluctuation of gluons. In SU(2) lattice case, the extraction is easy to be done without any approximation, because of the SU(2) property.
 With this extraction, the SU(2) lattice simulation suggests that the off-diagonal gluons do not propagate in the infrared region due to the effective mass $M_{\rm off} \simeq 1.2 {\rm GeV}$, while the diagonal gluon widely propagates \cite{AS98}.
 
 The aim of this paper is to propose a method to extract the gluons from the link-variable directly and generally in SU(3) lattice QCD, and to investigate the gluon propagators in the MA gauge.

\section{SU(3) Formalism and gluon propagators in MA gauge with U(1)$_3\times$U(1)$_8$ Landau gauge}
\label{3-1}
Using the SU(3) lattice QCD, we calculate the gluon propagators in the MA gauge with the U(1)$_3\times$U(1)$_8$ Landau gauge fixing. In the MA gauge, to investigate the gluon propagators, we use the gluon fields extracted directly from the link-variables \cite{FN04}.
Here, we analytically extract the gluon field $A_\mu (x)$ 
from each link-variable $U_\mu(x)$ as follows:
\begin{eqnarray}
A_\mu (x) = \frac{1}{iag}{\rm Ln} U _\mu (x)= \frac{1}{iag}
\Omega  ^\dagger _\mu (x){\rm Ln} U_\mu ^d (x) \Omega _\mu (x),
\end{eqnarray}
where ${\rm Ln}$ is the natural logarithm defined on complex numbers, 
$U^d_\mu (x)$ is the diagonalized unitary matrix, 
and $\Omega _\mu (x)$ is the diagonalization unitary matrix of eigenvectors. 

The MA gauge fixing is performed by the maximization of 
\begin{eqnarray}
R_{\rm MA}
	&\equiv&
	 \sum_x \sum^4_{\mu =1} 
	{\rm tr} \left[ U_\mu (x) \vec{H} U_\mu^{\dagger}(x) \vec{H}  \right],
						\label{eqn:Rma}
\end{eqnarray}
where $\vec{H} =(T_3,T_8)$ is the Cartan generator. In this gauge fixing, there remains U(1)$_3\times$U(1)$_8$ gauge symmetry. In order to study the gluon propagators, we fix the residual gauge. After the Cartan decomposition for the SU(3) link-variables as $U_\mu (x) \equiv M_\mu (x) u_\mu (x)$ with $u_\mu(x) \equiv e^{i(\theta^3(x)T ^3 + \theta ^8 (x) T ^8 )}\in$U(1)$_3\times$U(1)$_8$ and $M_\mu (x)= e^{i\sum_{a\ne 3,8} \theta ^a (x)T^a} \in$SU(3)$/$U(1)$_3\times$U(1)$_8$, the residual gauge fixing is performed 
by the maximization of 
\begin{eqnarray}
R_{\rm U(1)L}\equiv \sum_x \sum_{\mu=1}^4 {\rm Re}\ {\rm tr}[u_\mu(x)].
											\label{eqn:defU1Landau}
\end{eqnarray}
At $\beta$=5.7, 5.8 and 6.0, the lattice spacings $a$ are estimated as $a \simeq 0.186{\rm fm},0.152{\rm fm}$ and $0.104{\rm fm}$, respectively, which lead to the string tension $\sigma \simeq 0.89 \mathrm{GeV/fm} $ in the inter-quark potential \cite{IS09}.

After gauge fixing completely, we study the Euclidean scalar combination of 
the diagonal (Abelian) gluon propagator as
\begin{eqnarray}
G_{\mu\mu}^{\rm Abel}(r) \equiv \frac{1}{2} \sum_{a= 3,8} \left< A_\mu^a(x)A_\mu^a(y)\right>,
				\label{eqn:AAf002}
\end{eqnarray}
and that of the off-diagonal gluon propagator as
\begin{eqnarray}
G_{\mu\mu}^{\rm off}(r) \equiv
\frac{1}{6} \sum_{a\neq 3,8} \left< A_\mu^a(x)A_\mu^a(y)\right>.
				\label{eqn:AAf003}
\end{eqnarray}
These are expressed as the function of the four-dimensional Euclidean distance $r\equiv \sqrt{(x_\mu -y_\mu)^2}$. 
When we consider the renormalization, 
these propagators are multiplied by an $r$-independent constant, 
according to the renormalized gluon fields 
obtained by multiplying a constant renormalization factor.

The Monte Carlo simulation is performed  with the standard plaquette action on the $16^4$ lattice with $\beta$ =5.7, 5.8 and 6.0 at the quenched level. All measurements are done every 500 sweeps after a thermalization of 10,000 sweeps using the pseudo heat-bath algorithm. We prepare 50 gauge configurations for the calculation at each $\beta$. The error is estimated with the jackknife analysis.

\begin{figure}[h]
\begin{center}
\includegraphics[scale=0.6]{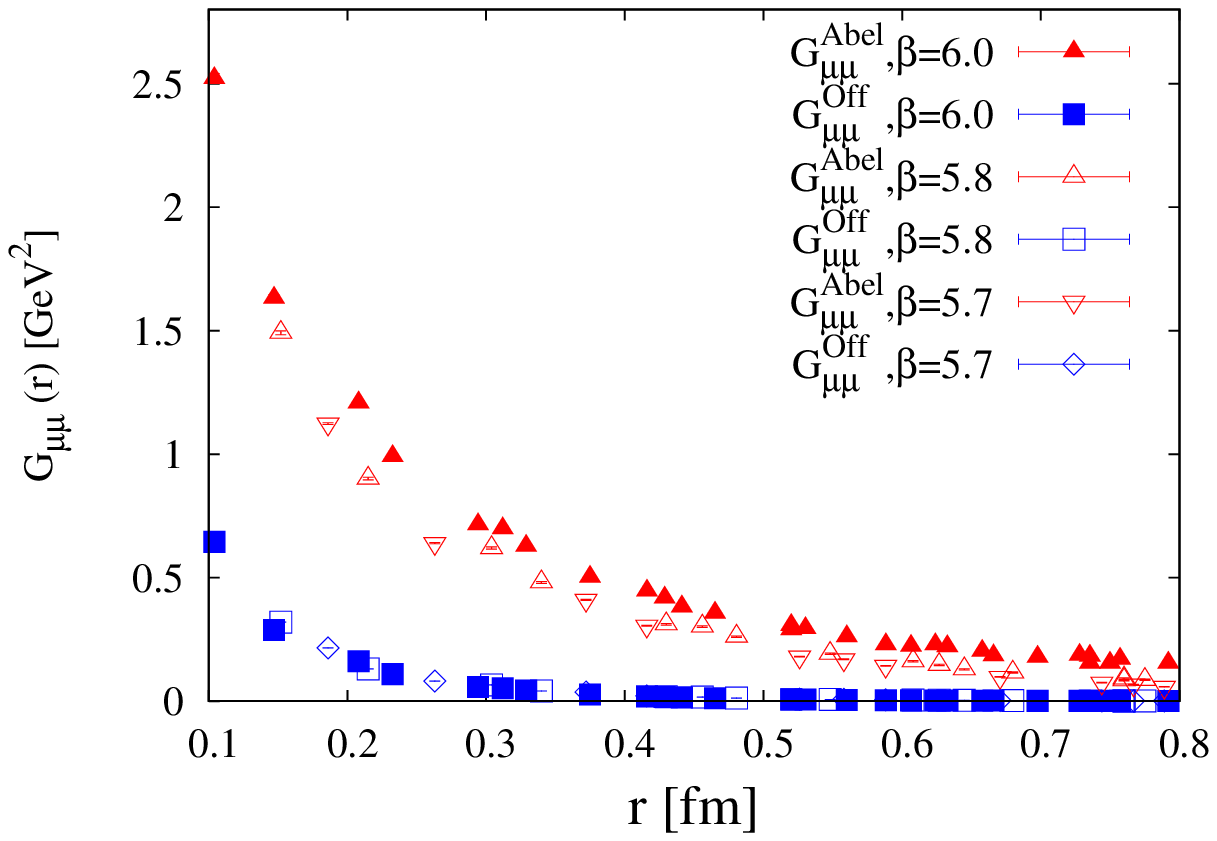}
\includegraphics[scale=0.6]{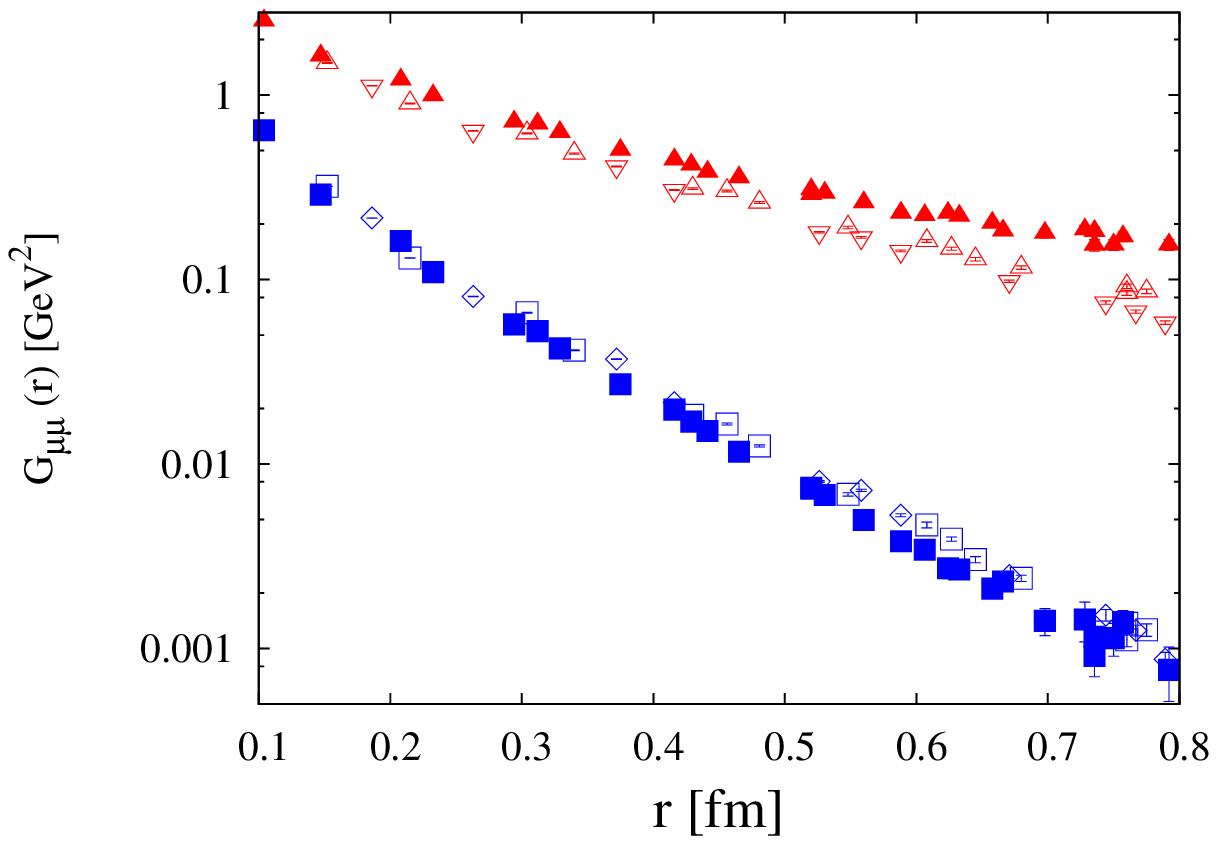}
\caption{
The SU(3) lattice QCD results of $G_{\mu\mu}^{\rm Abel}(r)$ and $G_{\mu\mu}^{\rm off}(r)$ (top), and their logarithmic plots (bottom) as the function of $r\equiv \sqrt{(x_\mu -y_\mu)^2}$ in the MA gauge with the U(1)$_3\times$U(1)$_8$ Landau gauge fixing in the physical unit. The Monte Carlo simulation is performed on the $16^4$ lattice with $\beta$ = 5.7, 5.8 and 6.0. The diagonal-gluon propagator $G_{\mu\mu}^{\rm Abel}(r)$ takes a large value even at the long distance. On the other hand, the off-diagonal gluon propagator $G_{\mu\mu}^{\rm off}(r)$ rapidly decreases.
}
\label{Fig1}
\end{center}
\end{figure}

We show in Fig.\ref{Fig1} the lattice QCD result for the diagonal gluon propagator $G_{\mu\mu}^{\rm Abel}(r)$ and the off-diagonal gluon propagator $G_{\mu\mu}^{\rm off}(r)$ in the MA gauge with the U(1)$_3\times$U(1)$_8$ Landau gauge fixing. In the MA gauge, $G_{\mu\mu}^{\rm Abel}(r)$ and 
$G_{\mu\mu}^{\rm off}(r)$ manifestly differ.
The diagonal-gluon propagator $G_{\mu\mu}^{\rm Abel}(r)$ 
takes a large value even at the long distance. 
In fact, the diagonal gluons $A_\mu^3,A_\mu^8$ in the MA gauge 
propagate over the long distance.
On the other hand, the off-diagonal gluon propagator 
$G_{\mu\mu}^{\rm off}(r)$ rapidly decreases and is negligible
for $r \gsim 0.4$ fm in comparison with $G_{\mu\mu}^{\rm Abel}(r)$. 
Then, the off-diagonal gluons $A_\mu^a~(a\neq 3,8)$ seem to propagate 
only within the short range as $r \lsim 0.4$ fm.
Thus, ``infrared abelian dominance" is found in the MA gauge.

\section{Estimation of off-diagonal gluon mass in MA gauge}
\label{3-2}
Next, we investigate the effective gluon mass.
We start from the Lagrangian of 
the free massive vector field $A_\mu$ with the mass $M \ne 0$ 
in the Proca formalism,
\begin{eqnarray}
{\cal L}&=& \frac{1}{4}(\partial_\mu A_\nu - \partial _\nu A_\mu)^2
	+\frac{1}{2}M^2A_\mu A_\mu,   \label{eqn:Lag} 
\end{eqnarray}
in the Euclidean metric. The scalar combination of the propagator $G_{\mu\mu}(r;M)$ can be expressed 
with the modified Bessel function $K_1(z)$ as 
\begin{eqnarray}
{G}_{\mu\mu}(r;M) &=& \left< A_\mu(x) A_\mu(y) \right> \nonumber \\
	&=&\int\frac{d^4 k}{(2\pi)^4} e^{i k \cdot (x-y)}
	 \frac{1}{k^2+M^2}
 	\left( 4+\frac{k^2}{M^2}\right) \nonumber\\
	&=&
	3\int\frac{d^4k}{(2\pi)^4} e^{i k \cdot (x-y)}	\frac{1}{k^2+M^2}    
	 + \frac{1}{M^2}{\delta^4(x-y)} \nonumber \\
	&=&
	\frac{3}{4\pi^2}\frac{M}{r}K_1(Mr)+\frac{1}{M^2}\delta^4(x-y).
					\label{eqn:prp02}
\end{eqnarray}
In the infrared region with large $Mr$, 
Eq. (\ref{eqn:prp02}) reduces to 
\begin{eqnarray}
G_{\mu\mu}(r;M)
	&\simeq&
	\frac{3\sqrt{M}}{2(2\pi)^{\frac{3}{2}}} \frac{e^{-Mr}}{r^\frac{3}{2}}, 
						\label{eqn:prp03} 
\end{eqnarray}
using the asymptotic expansion, 
\begin{eqnarray}
K_1(z) \simeq
	\sqrt{\frac{\pi}{2z}} e^{-z} 
	\sum^\infty_{n=0}
	\frac{\Gamma(\frac{3}{2}+n)}{n!\Gamma(\frac{3}{2}-n)} \frac{1}{(2z)^n}, 
						\label{eqn:prp03A}
\end{eqnarray}
for large ${\rm Re}~z$. 

\begin{figure}[h]
\begin{center}
\includegraphics[scale=0.4]{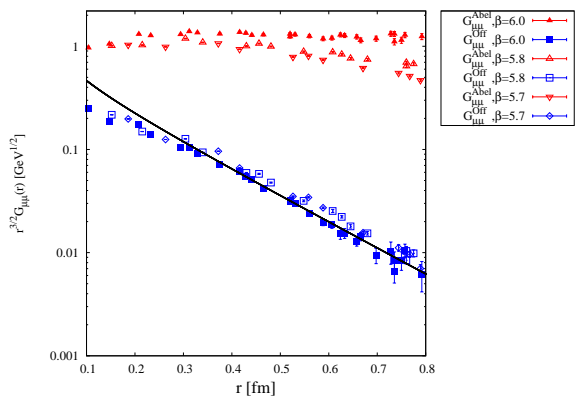}
\caption{
The logarithmic plot of $r^{3/2}G_{\mu\mu}^{\rm off} (r)$ and $r^{3/2}G_{\mu\mu}^{{\rm Abel}} (r)$ as the function of the four-dimensional Euclidean distance $r$ in the MA gauge with the U(1)$_3\times$U(1)$_8$ Landau gauge fixing, using the SU(3) lattice QCD with $16^4$ at $\beta$ = 5.7, 5.8 and 6.0. The solid line denotes the logarithmic plot of $r^{3/2}G_{\mu\mu}(r) \sim r^{1/2}K_1 (Mr)$ in the Proca formalism.
}
\label{Fig2}
\end{center}
\end{figure}

In Fig.\ref{Fig2}, we show the logarithmic plot of $r^{3/2}G_{\mu\mu}^{\rm off} (r)$ and $r^{3/2}G_{\mu\mu}^{\rm Abel} (r)$ as the function of the four-dimensional Euclidean distance $r$ in the MA gauge with the U(1)$_3\times$U(1)$_8$ Landau gauge fixing. From the linear slope on $r^{3/2}G_{\mu\mu}^{\rm off} (r)$, the effective off-diagonal gluon mass $M_{\rm off}$ is estimated. 
Note that the gluon-field renormalization does not affect the gluon mass 
estimate, since it gives only an overall constant factor for the propagator. 
We summarize in Table~I the effective off-diagonal gluon mass $M_{\rm eff}$ 
obtained from the slope analysis in the range of $r=0.3-0.8{\rm fm}$ 
at $\beta$ =5.7, 5.8 and 6.0.
The off-diagonal gluons seem to have a large effective mass of 
$M_{\rm off} \simeq 1.1-1.2 {\rm GeV}$.
This result approximately coincides with SU(2) lattice calculation \cite{AS98}.

Also for the diagonal gluon, we try to estimate its effective mass $M_{\rm diag}$, although its propagator largely depends on $\beta$, i.e., the volume or the spacing, as is indicated in Fig.\ref{Fig2}. We estimate the diagonal gluon mass $M_{\rm diag}$ from the slope analysis in the range of $r=0.3-0.8{\rm fm}$ at each $\beta$, and add the result in Table~I. In any case, the diagonal gluon seems to have a small effective mass of $M_{\rm diag} \simeq 0.1-0.3 {\rm GeV}$.
For the definite argument on $G_{\mu\mu}^{\rm Abel}(r)$ and the diagonal gluon mass, more careful analysis with a large-volume lattice would be needed.
\begin{table}[h]
\caption{Summary table of conditions and results in SU(3) lattice QCD. 
The off-diagonal gluon mass $M_{\rm off}$ is 
estimated from the slope analysis of $r^{3/2}G_{\mu\mu}^{\rm off} (r)$ for $r=0.3-0.8{\rm fm}$ at each $\beta$.
In the MA gauge, the off-diagonal gluons seem to have a large effective mass 
of $M_{\rm off} \simeq 1.1-1.2\mathrm{GeV}$.
The best-fit mass parameter $m_{\rm off}$ is also listed at each $\beta$: 
$G_{\mu\mu}^{\rm off} (r)$ in the range of $r=0.1-0.8 {\rm fm}$ is well described with the four-dimensional Euclidean Yukawa function $\sim e^{-m_{\rm off}r}/r$ with $m_{\rm off} \simeq 1.3 - 1.4\mathrm{GeV}$. 
We add the diagonal gluon effective mass $M_{\rm diag}$ at each $\beta$, 
estimated in a similar manner to $M_{\rm off}$.}
\begin{center} 
\begin{tabular}{cccccc}
\hline
\hline
  lattice size   & $\beta$     & $a[{\rm fm}]$ &   $M_{\rm off} [{\rm GeV}] $  & $m_{\rm off} [{\rm GeV}] $ &   $M_{\rm diag} [{\rm GeV}] $ \\
\hline
                 &    ~~~5.7~~~  & 0.186 &  1.2  & 1.3 & 0.3\\
$16^4$                 &    ~~~5.8~~~  & 0.152 &  1.1 & 1.3 & 0.2\\
                 &    ~~~6.0~~~     &  0.104 & 1.1 & 1.4 & 0.1\\
\hline
\hline
\end{tabular}
\end{center} 
\end{table}

Finally in this section, we discuss the relation 
between infrared abelian dominance and 
the off-diagonal gluon mass.
Due to the large effective mass $M_{\rm off} $, 
the off-diagonal gluon propagation is restricted within about
$M_{\rm off}^{-1} \simeq 0.2$ fm in the MA gauge.
Therefore, at the infrared scale as $r \gg 0.2$ fm,
the off-diagonal gluons $A_\mu^a~(a\neq 3,8)$ cannot mediate the long-range force like the massive weak bosons in the Weinberg-Salam model, 
and only the diagonal gluons $A_\mu^3,~A_\mu^8$ can mediate  
the long-range interaction in the MA gauge.
In fact, in the MA gauge, the off-diagonal gluons are expected to be 
inactive due to the large mass $M_{\rm off}$ in the infrared region 
in comparison with the diagonal gluons. 
Then, infrared abelian dominance holds for $r \gg M^{-1}_{\rm off}$. 

\section{Analysis of functional form of off-diagonal gluon propagator in MA gauge}
\label{3-3}
In this section, we investigate the functional form of the off-diagonal gluon propagator in the MA gauge in SU(3) lattice QCD. In the previous section, we compare the gluon propagator with the massive vector boson propagator and estimate the gluon mass. In fact, as shown in Fig.\ref{Fig2}, the gluon propagator would not be described by a simple massive propagator Eq. (\ref{eqn:prp02}) in whole region of $r=0.1-0.8 {\rm fm}$.

There is the similar situation in the Landau gauge \cite{IS09}. The functional form of the gluon propagator cannot be described by the propagator of the Proca formalism in whole region of $r=0.1-1.0 {\rm fm}$. The appropriate form is the four-dimensional Euclidean Yukawa-type function $\exp(-mr)/r$ with a mass parameter $m$.

\begin{figure}[h]
\begin{center}
\includegraphics[scale=0.4]{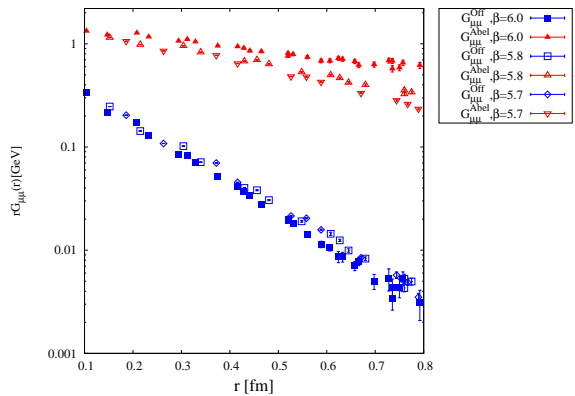}
\caption{
The logarithmic plot of 
$rG_{\mu\mu}^{\rm off} (r)$ and $rG_{\mu\mu}^{{\rm Abel}} (r)$ 
as the function of the four-dimensional Euclidean distance $r$ 
in the MA gauge with the U(1)$_3\times$U(1)$_8$ Landau gauge fixing, 
using the SU(3) lattice QCD with $16^4$ at $\beta$=5.7, 5.8 and 6.0. 
For $rG_{\mu\mu}^{\rm off} (r)$, 
the approximate linear correlation is found.}
\label{Fig3}
\end{center}
\end{figure}

In the same way, in the MA gauge, we also compare the gluon propagator with the four-dimensional Euclidean Yukawa function.
 In Fig.\ref{Fig3}, we show the logarithmic plot of $rG_{\mu\mu}^{\rm off} (r)$ and $rG_{\mu\mu}^{\rm Abel} (r)$ as the function of the distance $r$ in the MA gauge with the U(1)$_3\times$U(1)$_8$ Landau gauge fixing.
Note that, in the whole region of $r=0.1-0.8 {\rm fm}$, 
the logarithmic plot of $rG_{\mu\mu}^{\rm off} (r)$ is almost linear, 
and therefore the off-diagonal gluon propagator is well expressed 
by the four-dimensional Euclidean Yukawa function,
\begin{eqnarray}
G_{\mu\mu}^{\rm off} (r) \simeq A\frac{e^{-m_{\rm off} r}}{r},
\end{eqnarray}
with a mass parameter $m_{\rm off}$ and a dimensionless constant $A$.
The best-fit mass parameter $m_{\rm off}$ is given in Table~I at each $\beta$ = 5.7, 5.8 and 6.0. 

We comment on the four-dimensional Euclidean Yukawa-type propagator \cite{IS09}. If the functional form of the off-diagonal gluon is well discribed by the four-dimensional Yukawa function, we analytically calculate the off-diagonal zero-spatial-momentum propagator,
\begin{eqnarray}
D_0^{\rm off}  (t)\equiv \int d^3x G_{\mu\mu}^{\rm off} (r),
\end{eqnarray}
and obtain the spectral function $\rho (\omega)$ by the inverse Laplace transformation. Also in the MA gauge, the spectral function is found to have the negative-value region, as in the Landau gauge \cite{MO87,Bo04,IS09}.

\section{Summary and Concluding Remarks}
\label{4}
We have performed the first study on the gluon propagators in the MA gauge with the U(1)$_3\times$U(1)$_8$ Landau gauge fixing using the SU(3) lattice QCD. 
 To investigate the gluon propagators in the MA gauge, we have considered to derive the gluon fields from the SU(3) link-variables. In this method,
the gauge fields have been extracted by diagonalizing the link-variables and taking the logarithm. Owing to this method, any quantity expressed by the gluon fields can be calculated directly from link-variables, even if $|agA_\mu (x) | \ll 1$ does not satisfy.
As one of the general merits of this method,
we can directly check the correspondence between gluon fields and
the continuum gauge fixing in arbitrary lattice gauge fixing
performed with link-variables. In principle, with this method, continuum gauge fixing with gluon fields
can be also performed directly.

With this method, we have measured the Euclidean scalar combinations of the propagators $G_{\mu\mu} (r)$ for the diagonal and the off-diagonal gluons, and found the infrared Abelian dominance. 
The Monte Carlo simulation is performed on the $16^4$ lattice with $\beta$ =5.7, 5.8 and 6.0 at the quenched level.
We have found that 
the off-diagonal gluons behave as massive vector bosons 
with the effective mass $M_{\rm off} \simeq 1.1-1.2$ GeV for $r =0.3 -0.8$ fm. The effective gluon mass has been estimated from the linear fitting analysis of the logarithmic plot of $r^{3/2}G_{\mu\mu} ^{\rm off}(r)$. Due to the large value, the finite-size effect for the off-diagonal gluon mass is expected to be ignored. The large gluon mass shows that the off-diagonal gluons cannot mediate the interaction over the large distance as $r \gg M_{\rm off}^{-1}$, and such an infrared inactivity of the off-diagonal gluons would lead infrared Abelian dominance in the MA gauge.

On the other hand, from the behavior of the diagonal gluon propagator $G_{\mu\mu} ^{\rm Abel}(r)$ and $r^{3/2}G_{\mu\mu} ^{\rm Abel}(r)$, the diagonal gluons seem to behave as light vector bosons with the effective mass $M_{\rm diag} \simeq 0.1-0.3$ GeV for $r =0.3 -0.8$ fm, 
although careful analysis with a large-volume lattice is needed 
for more definite argument on $G_{\mu\mu}^{\rm Abel}(r)$.


Finally, we have also investigated the functional form of the propagator 
in the MA gauge. We show that the off-diagonal gluon propagator 
$G_{\mu\mu}^{\rm off}(r)$ is well described by 
the four-dimensional Euclidean Yukawa-type form with the mass parameter 
$m_{\rm off} \simeq 1.3-1.4$ GeV in the region of $r=0.1-0.8$ fm.
This indicates that the spectral function $\rho (\omega)$ of the off-diagonal gluons in the MA gauge has the negative-value region as in the Landau gauge.

On the other hand, the functional form of the diagonal gluon propagator seems to be the four-dimensional Euclidean Yukawa function with the lighter mass parameter. However, to discuss the functional form clearly, the finite size effect is to be checked carefully just like the estimation of the diagonal effective gluon mass. 

In this study, we investigate the off-diagonal gluon propagator. To be strict, the off-diagonal gluon propagator consists of two scalar functions corresponding to longitudinal and transverse components. Therefore, we will investigate each effective mass and the functional form of these components.

\section*{Acknowledgements}
The authors are deeply grateful to Dr. Hideaki Iida for useful discussions.  
This work is supported in part by the Grant for Scientific Research 
[(C) No.~23540306, Priority Areas ``New Hadrons'' (E01:21105006)], Grant-in-Aid for JSPS Fellows (No.23-752, 24-1458)
from the Ministry of Education, Culture, Science and Technology 
(MEXT) of Japan, and the Global COE Program, 
``The Next Generation of Physics, Spun from Universality and Emergence".
The lattice QCD calculations are done on NEC SX-8R at Osaka University.

\end{document}